\documentclass[amsmath,amssymb,aps,prc,twocolumn,floatfix,showpacs]{revtex4-1}
\usepackage{graphicx}
\usepackage{dcolumn}
\usepackage{bm}
\usepackage{color}
\begin{document}


\title{No-Core MCSM calculation for $^{10}$Be and $^{12}$Be low-lying spectra}

\author{Lang Liu}
\email[]{liulang@pku.edu.cn}
\affiliation{Department of Physics, University of Tokyo, Hongo, Tokyo 113-0033, Japan  \\
State key Laboratory of Nuclear Physics and Technology, School of Physics, Peking University, Beijing 100871, P. R. China
}

\author{Takaharu Otsuka}
\affiliation{Department of Physics, University of Tokyo, Hongo, Tokyo 113-0033, Japan  \\
Center for Nuclear Study, University of Tokyo, Hongo, Tokyo 113-0033, Japan  \\
National Superconducting Cyclotron Laboratory, Michigan State University, East Lansing, Michigan, 48824, USA}

\author{Noritaka Shimizu}
\affiliation{Department of Physics, University of Tokyo, Hongo, Tokyo 113-0033, Japan}

\author{Yutaka Utsuno}
\affiliation{Japan Atomic Energy Agency, Tokai, Ibaraki, 319-1195 Japan}

\author{Robert Roth}
\affiliation{Institut f\"{u}r Kernphysik, Technische Universit\"{a}t Darmstadt, D-64289 Darmstadt, Germany}
\date{\today}

\begin{abstract}

The low-lying excited states of $^{10}$Be and $^{12}$Be are investigated 
within a no-core Monte Carlo Shell Model (MCSM) framework employing 
a realistic potential obtained via the Unitary Correlation Operator
Method. The excitation energies of the 2$^+_1$ and 2$^+_2$ states and 
the B(E2;$\,2^+_{1,2}\rightarrow$ 0$^+_{g.s.}$) for $^{10}$Be in the 
MCSM with a standard treatment of spurious center-of-mass motion show 
good agreement with experimental data. Some properties of 
low-lying states of $^{10}$Be are studied in terms of 
quadrupole moments, E2 transitions and single-particle occupation
numbers. 
The E2 transition probability of $^{10}$C, the mirror nucleus of 
$^{10}$Be, is also presented with a good agreement to experiment. 
The triaxial deformation of $^{10}$Be and $^{10}$C is discussed 
in terms of the B(E2) values. 
The removal of the spurious center-of-mass motion 
affects differently on various states: for instance, negligible 
effects on the 2$^+_1$ and 2$^+_2$ levels of $^{10}$Be, 
while significant and favorable shift for the 1$^-_1$ level.
It is suggested that the description of $^{12}$Be needs a larger
model space as well as some other higher excited states of $^{10}$Be, 
as an indicator that these are dominated by intruder configurations.

\end{abstract}


\maketitle

\section{\label{sec1}Introduction}

One of the major goals in nuclear physics is to understand the structure
and reactions of nuclei starting from realistic nuclear
interactions. Besides the challenge of solving the nuclear many-body
problem, this endeavor is complicated by the fact that our understanding
of the nuclear force is not complete yet. At present, there are two ways
to construct an accurate representation of nuclear force. One can
construct a two-body potential phenomenologically by fitting
experimental data on nucleon-nucleon (\textit{NN}) scattering, as it is
done in the Argonne V18 potential~\cite{Wiringa1995}, the CD-Bonn
potential~\cite{Machleidt2001} and the Nijmegen
potentials~\cite{Stoks1994}. Alternatively, consistent two- and
many-body interactions can be constructed in the framework of chiral
effective field theory using the symmetries and the effective degrees of
freedom of low-energy QCD as a guiding principle. The chiral N$^{3}$LO potential is such an accurate charge-dependent nucleon-nucleon potential constructed at fourth order of chiral perturbation theory~\cite{Entem2003,Epelbaum2006654,Machleidt2011}. By using these realistic nuclear interactions, \textit{ab initio} nuclear many-body calculations have been performed in the last decade. In Green's Function Monte Carlo (GFMC) calculations the exact ground-state wave function is calculated by treating the many-body Green's functions in a Monte Carlo approach~\cite{Pieper2001,Pieper2002,Pieper2005}. The GFMC calculations of light nuclei up to $^{12}$C with the Argonne interaction reproduce the experimental nuclear binding energies and radii as well as the spectra. Another \textit{ab initio} approach for nuclei up to A=14 is the No-Core Shell Model (NCSM)~\cite{Navratil2000,Caurier2002,PetrNavratil2009}. All nucleons are treated in a large number of shell-model basis, providing similarly successful description of light nuclei.

However, the straightforward application of those realistic interactions in nuclear many-body calculations is still difficult due to the strong short-range repulsion which generates strong correlations in the nuclear many-body state. The Unitary Correlation Operator Method (UCOM) is one of the methods to tackle this problem by introducing a unitary transformation such that the transformed many-body states contain the information on the dominant correlations in nuclear many-body system~\cite{Feldmeier1998,Neff2003,Roth2010}. In the UCOM approach two unitary transformation operators are defined, a central correlation operator and a tensor correlation operator, which correspond to two most important correlations: the central correlations induced by the strong short-range repulsion and the tensor correlations, respectively. Through a unitary transformation of the Hamiltonian, a soft phase-shift equivalent two-nucleon interaction can be obtained. This UCOM potential can be used in various kinds of many-body calculations, such as no-core shell model calculations~\cite{Roth2007a,Roth2007,Roth2008b,Roth2009}.

In the shell model calculations, the direct diagonalization of the
Hamiltonian matrix in the full valence-nucleon Hilbert space is
difficult, as the dimension of such a space becomes larger and larger
when one moves from light nuclei to heavier nuclei. Much effort of
truncation frameworks to full shell-model calculation has been directed,
e.g. in Refs~\cite{Roth2007a,Puddu2006,Puddu2007}.  As another way to
overcome this difficulty, the  stochastic approaches have been
introduced. Among them, the Shell Model Monte Carlo (SMMC) method has
been successfully proposed~\cite{Koonin1997}. Nevertheless, the SMMC is
basically suitable for the ground state and thermal properties, and
suffers from the so-called ``sign problem''. As a completely different
approach, the Quantum Monte Carlo Diagonalization (QMCD) method has been
proposed for solving quantum many-body systems with a two-body
interaction~\cite{Honma1995,Mizusaki1996,Honma1996,Otsuka1998}. The QMCD
can describe not only the ground state but also excited states,
including their energies, wave functions and hence transition matrix
elements.Thus, on the basis of the QMCD method, the Monte Carlo Shell
Model (MCSM) has been introduced~\cite{Otsuka2001} for nuclear shell
model calculations
~\cite{Mizusaki1999,Utsuno1999,Shimizu2001,Mizusaki2001,Utsuno2001}.
An extrapolation method in the Monte Carlo Shell Model has been proposed
very recently~\cite{Shimizu2010}. The applicability of the MCSM to a
system beyond the current limit of exact diagonalization is shown for
the $pf+g_{9/2}$-shell calculation by assuming a core in their work. It
is then of a certain importance and interest to apply the MCSM to
\textit{ab initio} calculations of light nuclei. As the MCSM has never
been used in \textit{ab initio} calculations, we start with
straightforward calculations by taking conventional MCSM method and code
which have been used for many shell-model calculations for medium-mass
nuclei.  We shall present, in this paper, how such \textit{ab initio}
calculations work. We note that the MCSM method is being revised in
parallel, and outlines of such revisions and future directions can be
found in Refs.~\cite{T.Abe2011,T.Abe}. The results to be shown in this
paper will play a key role in judging as to whether one should 
move ahead to more systematic calculations with the revised method or not.

The MCSM calculation is performed without a core to make it \textit{ab
initio}. In Section 2, we will introduce the theoretical framework of
the MCSM and explain the general procedure of the Monte Carlo Shell
Model method. $^{4}$He, which is investigated in the framework of the
shell model and the MCSM by using the UCOM potential, is discussed in
Section 3 as the numerical check. Study of structure and low lying
spectra for Beryllium isotopes appears in Section 4. In Section 5, the
conclusion with a summary and description of future direction for 
research in this field is given.

\section{\label{sec2}Monte Carlo Shell Model calculation}

The main idea of the MCSM is to diagonalize the Hamiltonian in a
subspace spanned by the MCSM basis states, which are generated in 
a stochastic way.

We begin with the imaginary-time evolution operator
\begin{eqnarray}\label{eq3_1}
e^{-\beta H},
\end{eqnarray}
where $H$ is a given general (time-independent) Hamiltonian and 
$\beta\propto T^{-1}$ is a real number with $T$ being analogous to a 
temperature. If this operator in Eq.~(\ref{eq3_1}) acts on a state 
$|\Psi^{(0)}\rangle $, one obtains
\begin{eqnarray}\label{eq3_2}
e^{-\beta H} |\Psi^{(0)}\rangle\,=\,\sum\limits_{i}e^{-\beta E_{i}}c_{i}|\psi_{i}\rangle,
\end{eqnarray}
where $E_{i}$ is the $i$-th eigenvalue of \textit{H}, 
$|\psi_{i}\rangle $ is the corresponding eigenstate and $c_{i}$ its 
amplitude in the initial state:
\begin{eqnarray}\label{eq3_3}
|\Psi^{(0)}\rangle\,=\,\sum\limits_{i}c_{i}|\psi_{i}\rangle.
\end{eqnarray}
For $\beta$ large enough, only the ground and low-lying states survive. 
But the actual handling is very complicated for $H$ containing a 
two-body (or many-body) interaction.

The Hubbard-Stratonovich (HS) transformation
~\cite{Hubbard1959,Stratonovich1957} can be used to ease the difficulty 
mentioned above. We then move to the formula
\begin{eqnarray}\label{eq3_17}
|\Phi(\sigma)\rangle\,\varpropto\,e^{-\beta h(\sigma)}|\Psi^{(0)}\rangle,
\end{eqnarray}
where $h(\sigma)$ is a one-body Hamiltonian obtained through the HS-transformation and $\sigma$ is a set of random numbers (auxiliary fields). The right-hand-side of this relation can be interpreted as a means to generate all basis vectors needed for describing the ground state and the low-lying states. For different values of the random variable, $ \sigma $, one obtains different state vectors, $ |\Phi(\sigma)\rangle $, by Eq.~(\ref{eq3_17}). These vectors are labeled as candidate states and selected as MCSM basis by a procedure of energy comparison~\cite{Otsuka2001}.

During the MCSM generation of the basis vectors, symmetries,
e.g. rotational and parity symmetry, are restored before the
diagonalization as more basis vectors are included. All MCSM basis
states are projected onto good parity and angular momentum quantum
numbers by acting with the corresponding projection operators. We
diagonalize the Hamiltonian in a subspace spanned by those projected
basis vectors. The number of the MCSM basis states is referred to as the
MCSM dimension. The basis generation process for general cases is
outlined in Ref.~\cite{Otsuka2001}.

As more than one major shell is included in the MCSM calculation, the
spurious center-of-mass motion must be accounted for. The
Lawson's prescription is adopted to suppress the spurious center-of-mass
motion in good approximation for major shell
truncation~\cite{GLOECKNER1974}. The total Hamiltonian then consists of
intrinsic and center-of-mass parts as,
\begin{eqnarray}\label{eq3_31}
H'\,=\,H_{int.}+\beta_{c.m.} H_{c.m.},
\end{eqnarray}
where $ H_{int.} $ is the intrinsic Hamiltonian. The $ H_{c.m.} $ is defined by
\begin{eqnarray}\label{eq3_32}
H_{c.m.}\,=\,\frac{\mathbf{P}^2}{2AM}+\frac{1}{2}MA\omega^2\mathbf{R}^2-\frac{3}{2}\hbar\omega,
\end{eqnarray}
where $\mathbf{R}$ and $\mathbf{P}$ are the coordinate and momentum of
the center of mass, respectively. In general, by taking sufficiently
large values of $ \beta_{c.m.} $, spurious components are suppressed 
for the low-lying eigenstates of $H'$.

\section{\label{sec3}Results for $^{4}\text{He}$}

In this section we discuss the interactions and model spaces used for
the no-core MCSM and provide some benchmark calculations for the $^4$He
ground state. The model space of the MCSM is spanned by a harmonic
oscillator basis truncated with respect to the unperturbed
single-particle energies $e_{\rm max} = 2n+l$. We use UCOM-transformed
realistic two-nucleon interactions as input potential. In addition to
the standard UCOM interaction derived from the Argonne V18 potential,
which has been used in a series of applications in various many-body
methods \cite{Roth2010}, we adopt a new UCOM potential based on the
chiral N$^3$LO two-nucleon interaction of Entem and Machleidt
\cite{Entem2003,Machleidt2011}. These UCOM potentials are labeled as
$V_{\text{UCOM}}$(AV18) and $V_{\text{UCOM}}$(N$^3$LO), respectively. In
both cases the UCOM correlation functions are determined through an
energy minimization in the two-nucleon system with a constraint on the
range of the even spin-triplet tensor correlator \cite{Roth2010}. We
neglect Coulomb interaction in all of our calculations throughout this
work for simplicity.

As an example of the UCOM potential, we perform a straightforward shell 
model diagonalization within a harmonic oscillator basis without a
core. 
In this shell model calculation, we employ the NuShell code developed 
by B.A. Brown \textit{et al.}~\cite{Brown2007} and use 
the $V_{\text{UCOM}}$(N$^3$LO) potential.

\begin{figure}[tb]
  \centering
    \includegraphics[width=8cm,clip]{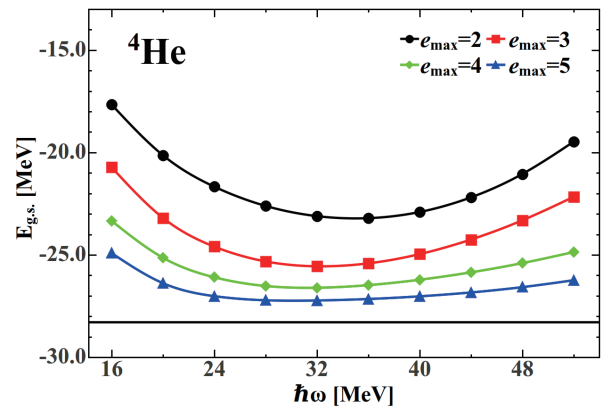}
  \caption{(Color online) The ground-state energy of $ ^{4}$He as a function of harmonic oscillator frequency $\hbar\omega$ in different model spaces ($e_{\rm max}$ from 2 to 5) calculated in NuShell using the $V_{\text{UCOM}}$(N$^3$LO) potential. The symbols correspond to $e_{\rm max}$ from 2 to 5. The straight line is experimental value.
  } \label{he4_sm_n3lo}
\end{figure}

The ground-state energy for $^{4}$He as a function of the harmonic 
oscillator frequency $\hbar\omega$ in various model spaces 
characterized by the oscillator basis cut off parameter $e_{\rm max}$, 
is shown in Fig.~\ref{he4_sm_n3lo}. 
The shell-model description of many-body correlations depends in 
general on the size of the model space. 
The experimental value is shown as a black line. The ground-state energy 
for small model spaces, e.g., $e_{\rm max}$=2, shows a sizable dependence on $\hbar\omega$. By increasing the size of the model space, the ground-state energy is lowered and dependence on $\hbar\omega$ is reduced. The ground-state energy varies by about 1 MeV for a range of oscillator frequencies $\hbar\omega$ from 24 MeV to 52 MeV. There is still about 1 MeV difference between the $e_{\rm max}$=5 result at $\hbar\omega$=32 MeV and the experimental ground-state energy. Evidently, still larger values of $e_{\rm max}$ are needed to reproduce the experimental binding energy of $^4$He, which the $V_{\text{UCOM}}$(N$^3$LO) interaction is approximately adjusted to. At present, the conventional shell model calculation is performed only up to $e_{\rm max}=5$ model space due to the limitation of Nushell code. However, the convergence is significantly better than for the bare realistic nucleon-nucleon potential. The comparison between bare interaction and the UCOM interaction in the shell model calculation for $^{4}$He can be found in Ref.~\cite{Roth2005a}.

\begin{figure}[tb]
\centering
 \includegraphics[width=8cm,clip]{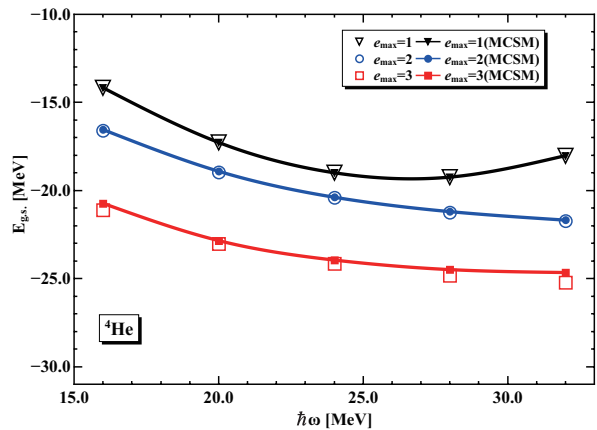}

 \caption{(Color online)  The ground-state energy of $^4$He as a
 function of the harmonic oscillator parameter $\hbar\omega$ using UCOM 
potential $V_{\text{UCOM}}$(AV18) in $e_{\rm max}$=1 (black), 2 (blue), 
and 3 (red) model space. The open symbols indicate results by the 
 conventional direct diagonalization method in the $m$-scheme. 
 The closed symbols indicate the MCSM results.
 }\label{he4_compare}
\end{figure}

We now compare the full shell-model and the MCSM results for the
ground-state energy of $^{4}$He using the $V_{\text{UCOM}}$(AV18)
potential. Figure~\ref{he4_compare} shows the ground-state energy for
$^{4}$He for both calculations as a function of the oscillator frequency
$\hbar\omega$ in small model spaces ($e_{\rm max}$=1, 2 and 3). The MCSM
results obtained with 32$\sim$50 MCSM dimensions are in reasonable 
agreement with the results from a full
diagonalization in these model spaces.

\begin{figure}[tb]
\centering
 \includegraphics[width=8cm,clip]{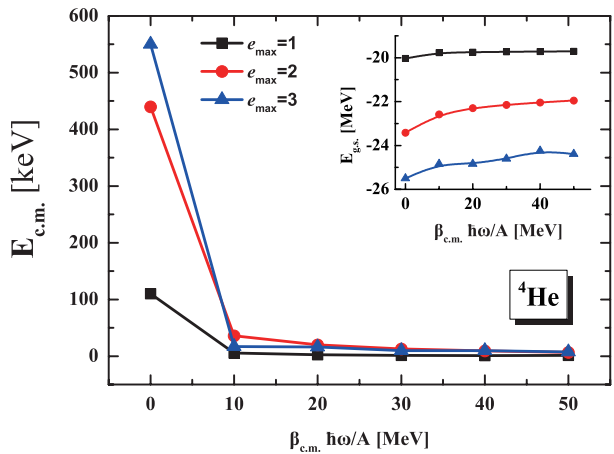}
 \caption{The center-of-mass motion energy and the ground-state energy 
 (inset) of $^{4}$He as a function of Lawson's prescription 
 parameter $\beta_{c.m.}$ defined in eq.~(\ref{eq3_31}). 
 The model spaces are $e_{\rm max}$=1, 2 and 3. 
 The oscillator parameter $\hbar\omega$ is $28$ MeV. 
 The potential $V_{\text{UCOM}}$(N$^3$LO) is used.
 }\label{he4_cm}
\end{figure}

The treatment of spurious center-of-mass motion of $^{4}$He is
illustrated in Fig.~\ref{he4_cm}. Figure~\ref{he4_cm} shows the
dependence of the expectation value of $H_{c.m.}$ and the ground-state
energy (inset) obtained with Lawson's prescription parameter
$\beta_{c.m.}$ in the $e_{\rm max}$=1, 2 and 3 model spaces. The
expectation value of $H_{c.m.}$ decreases rapidly and reaches a
converged small value. In this way, the spurious center-of-mass motion
can be suppressed to a large extent by choosing a suitable
$\beta_{c.m.}$ value.

\section{\label{sec4}low lying spectra of $^{10}\text{Be}$ and $^{12}\text{Be}$}

The $^{10}$Be nucleus is a good candidate for testing \textit{ab initio}
calculations employing realistic nuclear interactions, as there are
adequate experimental data both in the ground state and in the excited
states, e.g., excitation energies of two $J^{\pi}$=2$^{+}$, $T$=1 states
and the $B(E2)$ value of those states to the ground state. The AMD
calculations of Be isotopes~\cite{Kanada-En'yo1995}, the GFMC
approach~\cite{Pieper2001,Pieper2002} and the
NCSM~\cite{Caurier2002,Navratil2003} have been used to investigate
$p$-shell nuclei like $^{10}$Be and to reproduce features such as
binding energies and excitation spectra. This work is a new attempt to
investigate these states by applying the no-core MCSM with realistic
nuclear interactions. In this section, we present MCSM results for
$^{10}$Be and $^{12}$Be.

We discuss MCSM calculations using $V_{\text{UCOM}}$(N$^3$LO) potential in an $e_{\rm max}$=3 model space. For the beryllium isotopes, the ground-state energies exhibit a minimum for oscillator frequencies $\hbar\omega$ around $16.0$ MeV in the conventional shell model calculation~\cite{liuthesis}. We use bare charges, hence the electric quadrupole moment is equal to the proton quadrupole moment.

For a more precise investigation, we have to remove spurious components 
with respect to the center-of-mass motion, if they are mixed in
calculated eigenfunctions.  As discussed earlier, we use Lawson's
prescription with a suitably chosen $ \beta_{c.m.} $ in eq.~(\ref{eq3_31}). 
We shall use $\beta_{c.m.}\cdot\hbar\omega/A$ = 10 MeV hereafter, 
unless otherwise specified.  The same value has been taken in many
MCSM calculations (not of {\it ab initio} type), {\it e.g.}, 
\cite{Utsuno1999}.

\begin{figure}[tb]
\centering
 \includegraphics[width=8cm,clip]{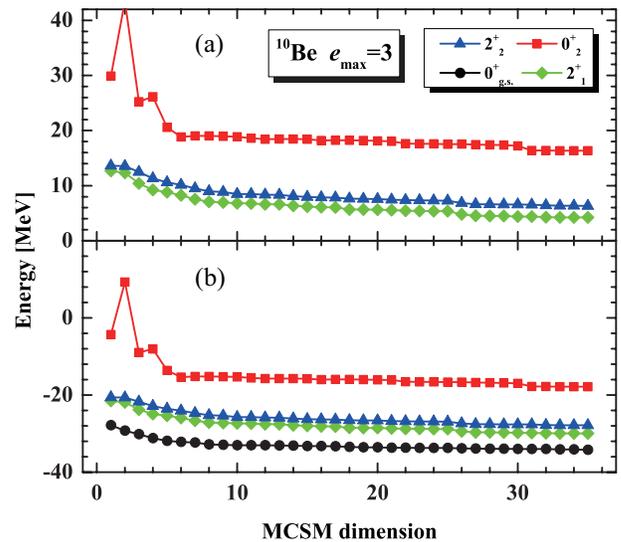}
 \caption{(Color online) (a) Excitation energies of the 2$^+_1$ state (green
 diamonds), the 2$^+_2$ state (blue triangles) and the 0$^+_2$ state (red
 squares) as functions of the MCSM dimension in the 
 $e_{\rm max}=3$ space. (b)
 Energies of the states shown in (a) as well as the 0$^+_1$ (ground) state
 (black circles). $\beta_{c.m.}\cdot\hbar\omega/A$ = 10 MeV is used to
 remove spurious components.
 }
 \label{be10_excited_dimension}
\end{figure}

The convergence of low-lying excitation energies as a function of the
MCSM dimension has to be examined, as our goal is to investigate the
excitation spectra. Figure~\ref{be10_excited_dimension} (a) shows the 
excitation energies of the 2$^{+}_{1}$, 2$ ^{+}_{2}$ and 0$^+_{2}$ 
states as functions of the MCSM dimension for the $e_{\rm max}$=3 
model space. If we evaluate the energy difference $\epsilon$ between 
results corresponding to the last two consecutive
MCSM dimensions, we obtain $\epsilon$=18 keV for 2$^+_1$, 
15 keV for 2$^+_2$ and 37 keV for 0$^+_2$. The relative accuracy of these
excitation energies is $\sim 0.3\%$ for 2$^+$ states and $\sim 0.7\%$
for 0$^+_2$ state. In the MCSM calculation, the diagonalization is
performed in a subspace comprised of 25 to 50 optimally generated 
basis states.
The size (dimension) of this subspace is quite small compared 
to that of the entire Hilbert space taken in the direct diagonalization 
in the conventional shell model. 
This advantage will be even more obvious for heavier nuclei by the fact 
that the full diagonalization in $e_{\rm max}=3$ is hardly feasible with 
other calculational techniques available presently.

Figure~\ref{be10_excited_dimension} (b) exhibits the energies of the  
0$^+_{1,2}$ and 2$^{+}_{1,2}$ states as functions of the MCSM
dimension.  One sees steady improvements of these energies, 
particularly for the dimension greater than 30.  The energies 
appear to become converged to a rather good extent.
Figure~\ref{be10_excited_dimension} (b) shows that the ground-state 
energy becomes about -35 MeV for dimensions large enough.
This is still far from the experimental value $\sim$ -65 MeV.
After the Coulomb correction $\sim$ 5 MeV, the difference is 
$\sim$ 35 MeV, which is in part due to the choice of 
the interaction where three-body forces are missing.  
The model space and the convergence also contribute to the 
discrepancy.  Such problems are important issues in present and 
future MCSM projects.
On the other hand, the present calculation appears to be rather 
reasonable for excitation energies as shown later, and 
we use the $V_{\text{UCOM}}$(N$^3$LO) potential in this first attempt.

\begin{figure*}[tb]
\centering
 \includegraphics[width=11cm,clip]{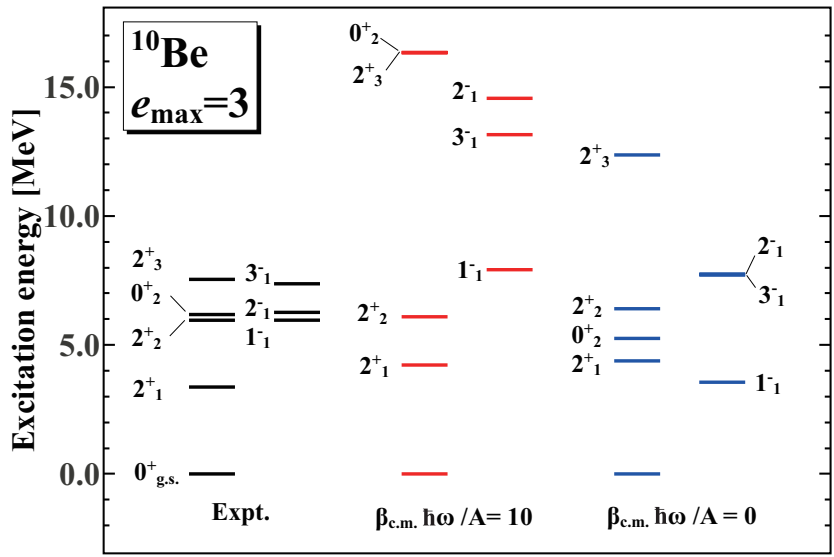}
 \caption{(Color online) Some low-lying spectra of $^{10}$Be in the 
 $e_{\rm max}=3$ model space.  Black bars indicate experimental levels.
 Red levels are theoretical results obtained with the suppression of 
 spurious center-of-mass
 motion ($\beta_{c.m.}\cdot\hbar\omega/A$ = 10 MeV). Blue levels are 
 obtained without removing the spurious center-of-mass motion.
 }
 \label{be10_excitation_beta}
\end{figure*}


\begin{figure}[tb]
\centering
 \includegraphics[width=9cm,clip]{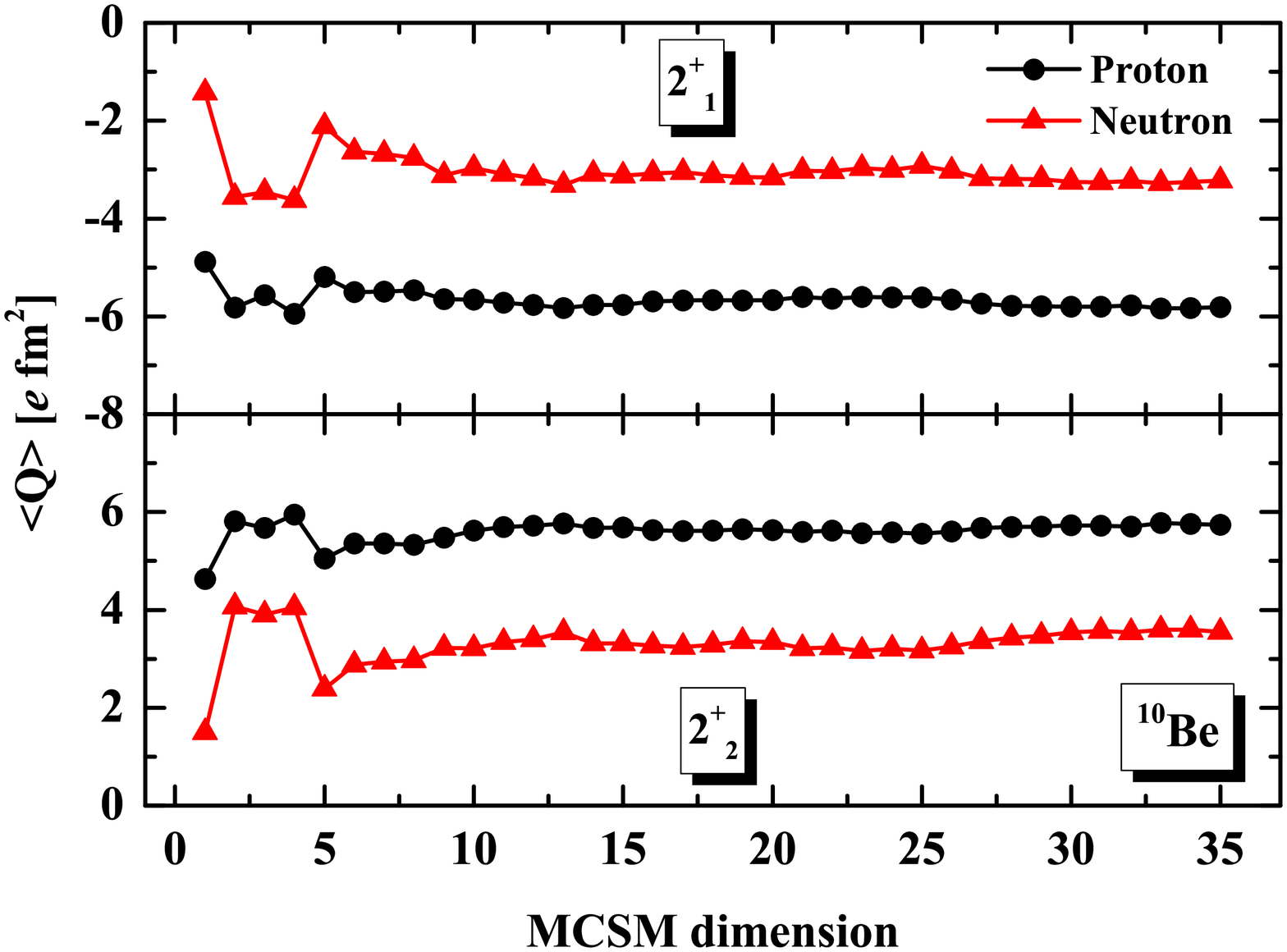}
 \caption{(Color online) Spectroscopic quadrupole moments for 
 protons (black circles) 
 and neutrons (red triangles) as functions of the MCSM dimension.
 Upper (lower) panel is for the 2$^+_1$ (2$^+_2$ ) state of $^{10}$Be.
 }
 \label{be10_q}
\end{figure}

We now discuss properties of the 0$^+_{1}$ and 2$^{+}_{1,2}$ states
of $^{10}$Be.
Figure~\ref{be10_excitation_beta} shows energy levels of these states.  
Some other low-lying states of $^{10}$Be are shown also, and will be 
discussed later. 
While the MCSM results are about 1 MeV and 0.6 MeV higher than 
the experimental values for 2$^+_1$ and 2$^+_2$, respectively, 
the basic patterns and scale are reproduced well by the MCSM 
calculation.  In particular, the low-lying 2$^{+}_{2}$ level is 
a characteristic indicator of triaxial deformation, as discussed
later.  

We now investigate these excited states in terms of the quadrupole
moments, E2 transitions and occupation probabilities. 
The quadrupole moments of protons and neutrons for the 2$^{+}_{1}$ and 
2$^{+}_{2}$ states of $^{10}$Be are shown, respectively, in 
Fig.~\ref{be10_q}. One finds that beyond MCSM 
dimension of 30, those quadrupole moments reach stable values. 
The nucleus $^{10}$Be has a negative quadrupole 
moment for the 2$^{+}_{1}$ state. In contrast, the 2$^{+}_{2}$ state 
shows a positive quadrupole moment. These features are also predicted 
in Ref.~\cite{Wiringa1995}. We note that the protons have stronger 
deformation than neutrons in both states of $^{10}$Be, because 
there are two valence protons and four valence neutrons in the $p$-shell 
in major configurations, and the former produce stronger deformation
than the latter. 


\begin{table*}[tb]
  \centering
\caption{B(E2) values (e$^{2}$~fm$^4$) of $^{10}$Be obtained by
 NCSM with CD-Bonn and CD-Bonn 2000
 potentials~\cite{Caurier2002,McCutchan_pri.}, GFMC with AV18 potential
 and AV18 plus different three-body forces~\cite{McCutchan_pri.}, 
 AMD~\cite{Kanada-En'yo1995}, present MCSM, and 
 experimental~\cite{McCutchan2009,McCutchan_pri.} values.}
\begin{tabular}{c|cc|ccc|c|c|c}
\hline\hline 
  & \multicolumn{2}{c|}{NCSM} & \multicolumn{3}{c|}{GFMC} & AMD  & MCSM & Expt. \\ 
\hline\hline
Quantity & CD-Bonn & CDB2k & AV18 & AV18+IL2 & AV18+IL7 &  &  &  \\ 
\hline 
 B(E2;$\,2^{+}_{1}\rightarrow 0^{+}_{g.s.}$) & 6.58 & 9.8(4) & 10.5(4) & 8.1(3) & 8.8(4) & 9.46 & 9.29 & 9.2(3) \\ \hline 
 B(E2;$\,2^{+}_{2}\rightarrow 0^{+}_{g.s.}$) & 0.13 & 0.2(2) & 3.4(2) & 3.3(2) & 1.8(1) &  & 0.32 & 0.11(2) \\ \hline 
\end{tabular}
\label{be2_10Be}
\end{table*}

\begin{table*}[tbh]
  \centering
  \caption{B(E2;$\,2^{+}_{1}\rightarrow 0^{+}_{g.s.}$), 
  B(E2;$\,2^{+}_{2}\rightarrow 0^{+}_{g.s.}$) and 
  B(E2;$\,2^{+}_{2}\rightarrow 2^{+}_{1}$) values (e$^2\,$fm$^4$) of 
  $^{10}$Be and B(E2;$\,2^{+}_{1}\rightarrow 0^{+}_{g.s.}$) of the
  mirror nucleus $^{10}$C obtained by the MCSM and the experimental 
  data~\cite{McCutchan2009,McCutchan_pri.}.}
  \begin{tabular}{c|ccc|ccc}
    \hline\hline
    & \multicolumn{3}{c|}{$^{10}$Be} & \multicolumn{3}{c|}{$^{10}$C} \\
    \hline\hline
  & B(E2;$\,2^{+}_{1}\rightarrow 0^{+}_{g.s.}$) &
	   B(E2;$\,2^{+}_{2}\rightarrow 0^{+}_{g.s.}$) &
   B(E2;$\,2^{+}_{2}\rightarrow 2^{+}_{1}$) &
   B(E2;$\,2^{+}_{1}\rightarrow 0^{+}_{g.s.}$) &
   B(E2;$\,2^{+}_{2}\rightarrow 0^{+}_{g.s.}$) &
   B(E2;$\,2^{+}_{2}\rightarrow 2^{+}_{1}$) \\
\hline
Exp. & 9.2(3) & 0.11(2) &  & 8.8(3) & & \\
MCSM & 9.29  & 0.32 & 3.28 & 9.30 & 2.15 & 12.81 \\
\hline\hline
  \end{tabular}
  \label{be2}
\end{table*}

The B(E2) values from the 2$^+_{1,2}$ states to the ground state 
of $^{10}$Be are shown in Table~\ref{be2_10Be}, in comparison to
results by NCSM~\cite{Caurier2002,McCutchan_pri.}, 
GFMC~\cite{McCutchan_pri.}, and AMD~\cite{Kanada-En'yo1995}.
The present result is rather similar to the CDB2k NCSM results
among those shown in this table.

Some B(E2) values are 
calculated also for the mirror nucleus, $^{10}$C, 
in the isospin formalism, as shown in Table~\ref{be2}.
This table indicates that  
MCSM value of B(E2;$\,2^{+}_{1}\rightarrow 0^{+}_{g.s.}$) 
appears to be in rather good agreement with the corresponding 
experimental data~\cite{McCutchan2009,McCutchan_pri.} 
for both $^{10}$Be and $^{10}$C.
This is of certain importance because from the viewpoint of the 
liquid-drop model, B(E2) value is proportional to $Z^2$, and 
thereby the value of $^{10}$C is expected to be larger than 
the corresponding one of $^{10}$Be, by a factor of $6^2/4^2$ 
in a naive expectation.  
While we take only bare charge ($e_p=e$ and $e_n=0$ with $e$ being 
the unit charge), we can still produce almost the same values of  
B(E2;$\,2^{+}_{1}\rightarrow 0^{+}_{g.s.}$) 
of $^{10}$Be and $^{10}$C.  This is because although there are 
two more protons in $^{10}$C than in $^{10}$Be, they do not 
necessarily increase quadrupole deformation, partly due to the 
0$p_{3/2}$ closed-shell formation.

We note that 
the B(E2;$\,2^{+}_{1}\rightarrow 0^{+}_{g.s.}$) value for $^{10}$C 
has been obtained by NCSM calculations as   
$5.702 ~e^2~$fm$^4$ ~\cite{Caurier2002} and $10\pm 2~e^2~$fm$^4$
~\cite{Forssen}.   A GFMC value has been reported as 
15.3 (1.4)~e$^2$~fm$^4$ ~\cite{McCutchan_pri.}. 
The present value, $9.3~e^2~$fm$^4$, appears to be the closest to
the observed value.

The MCSM value of the spectroscopic quadrupole moment of 
the 2$^+_1$ state of $^{10}$C is obtained also from Fig.~\ref{be10_q}
as 3.04 e~fm$^2$ by exchanging proton and neutron.

The nuclei $^{10}$C and $^{10}$Be belong to the same isospin multiplet 
of $T$=1. In the notation of Timmer~\cite{Timmer}, which makes direct 
use of the isospin formalism, one may write the E2 strength as
\begin{eqnarray}
	B(E2) &=& [(e_p+e_n)S+T_z(e_p-e_n)V]^2,
\end{eqnarray}
where the $e_p$ and $e_n$ are the effective charges being  
$e_p=e$ and $e_n=0$ in the present work. 
The reduced isoscalar and isovector matrix elements 
$S$ and $V$ must either be determined from experiment or be calculated 
with the help of suitable model wave functions. 
In Ref.~\cite{Alburger1969}, the B(E2) value is proportional to 
$[3.2+0.1\times T_z]^2$.  The B(E2) value of $^{10}$C should then be 
smaller than that in $^{10}$Be, as $T_z=-1$ for $^{10}$C and 
$T_z=1$ for $^{10}$Be.  

Assuming that the 0$^+_{1}$ and 2$^{+}_{1}$ states of $^{10}$Be 
belong to the same $K=0$ rotational band, the intrinsic quadrupole 
moment $Q_0$ can be evaluated from the 
B(E2;$\,0^{+}_{1} \rightarrow 2^{+}_{1}$) value and the spectroscopic 
quadrupole moment with the following relations~\cite{Bohr1998}
\begin{eqnarray}\label{quad}
	Q_0 &=& \dfrac{(I+1)(2I+3)}{3K^2-I(I+1)}Q, \\
	Q_0 &=& \left[ \frac{16\pi}{5}\cdot B(E2)\uparrow \right]^{1/2},
\label{quad2}	
\end{eqnarray}
where $Q$ is spectroscopic quadrupole moment, $K$ stands for the 
K quantum number, and $I$ is the angular momentum of a member of  
the rotational band. The intrinsic quadrupole moment evaluated by the 
spectroscopic quadrupole moment is 20.5 e\,fm$^2$, which is 
consistent to the one (21.6 e\,fm$^2$) extracted from the 
B(E2;$\,2^{+}_{1}\rightarrow 0^{+}_{1}$) value.   
This similarity seems to suggest an axially symmetric deformation
in the yrast band.  
On the other hand, the B(E2;$\,2^{+}_{2}\rightarrow 2^{+}_{1}$) is 
sizable, which hints at a notable triaxial deformation of $^{10}$Be. 
If $^{10}$Be has strictly axial deformation, this B(E2) value should be 
hindered, as the transition between $2^{+}_{2}$ state and $2^{+}_{1}$ 
state is forbidden by the selection rule of $K$ quantum number. 
The triaxiality leads to breaking of the $K$ selection rule.
For instance, B(E2;$\,2^{+}_{2}\rightarrow 0^{+}_{1}$)= 
$0.32~e^2~$fm$^4$ leads us to a triaxial deformation with 
$\gamma$ = 11.4$^\circ$ in the Davidov-Fillipov model~\cite{davydov1958}. 
Thus, the present results are of interest in view of nuclear shapes, 
although it may be an open question as to whether the classical picture 
of shapes can make sense in such light nuclei. It has been discussed 
in~\cite{Itagaki2002} that $^{10}$Be is triaxially deformed in a 
molecular-orbit calculation.

Table~\ref{be2} shows B(E2;$\,2^{+}_{2}\rightarrow 2^{+}_{1}$) and 
B(E2;$\,2^{+}_{2}\rightarrow 0^{+}_{1}$) of $^{10}$C too.  
These are considerably larger than the corresponding values of
$^{10}$Be.  At the first glance,   
the triaxiality appears to be more developed in $^{10}$C 
than in $^{10}$Be.  But one has to be careful, as this property 
holds for the proton part.
As the proton part and the neutron part are exchanged between
the mirror nuclei $^{10}$Be and $^{10}$C, it can be stated that
the part consisting of four protons in $^{10}$Be tends to be
deformed rather strongly in a prolate shape and the rest part 
(six neutrons) tends to be deformed in a triaxial
shape, and the situation is just reversed in $^{10}$C.  
The deformation here may be static, dynamic, in between, 
or even molecular ~\cite{Itagaki2002}. 
Such a difference between proton and neutron sectors 
is quite intriguing, and 
experimental investigations on these theoretical findings are of
much interest. 


Figure~\ref{be10_excitation_beta} shows the 2$^{+}_{1,2}$ levels
calculated in two ways (a) with center-of-mass motion suppression (default 
setup in this paper as stated already) and (b) without it for the
sake of comparison.
It is found that the suppression of the spurious center-of-mass 
motion keeps the excitation energies of the 2$^{+}_{1}$ and 2$^{+}_{2}$ 
states almost unchanged.  It seems that the suppression of the
c.m. motion is not so relevant to these states.

On the other hand, the c.m. motion suppression is essential to 
the 0$^{+}_{2}$, 1$^{-}_{1}$ and 2$^{+}_{3}$ states,  
as their excitation energies are raised by several $\sim$
10 MeV.


\begin{figure}[tb]
\centering
 \includegraphics[width=8cm,clip]{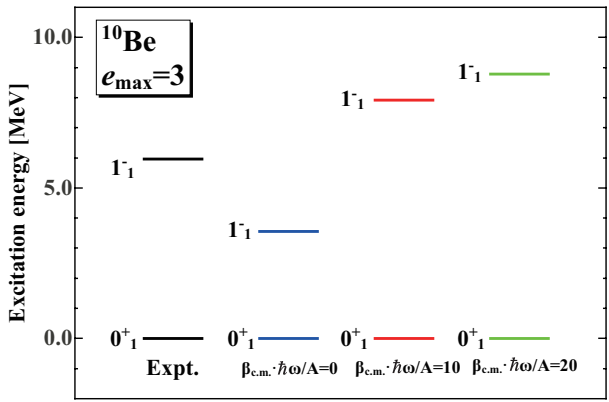}
 \caption{(Color online) The 1$^-_1$ excitation energy of $^{10}$Be 
 with different $\beta_{c.m.}$ values compared to the experimental 
 data (black bars). Blue, red and green bars indicate, respectively,  
 results with $\beta_{c.m.}\hbar\omega/A$= 0, 10 and 20~MeV.
 }
 \label{be10_1minus_beta}
\end{figure}
 
It is likely that the 1$^-_1$ state is sensitive to spurious center-of-mass 
contamination. We calculate the excitation energies of 1$^{-}_{1}$ state 
of $^{10}$Be with $\beta_{c.m.}\hbar\omega/A = 0, 10$ and
$20$~MeV. The levels are shown in Fig.~\ref{be10_1minus_beta}. 
When changing from 
$\beta_{c.m.}\hbar\omega/A =0$ to $10$~MeV, the excitation energy is 
increased by about 4 MeV. The result with 
$\beta_{c.m.}\hbar\omega/A = 20$~MeV is less than 1 MeV higher, 
presenting a hint at convergence, although it is slower than the 
2$^+_1$ state. 


The 0$^{+}_{2}$ and 2$^{+}_{3}$ levels in 
Fig.~\ref{be10_excitation_beta} are lying quite high 
compared to the experiment.
These states are expected to be intruder states with a large 
amount of $2p2h$ and higher excitations from the $p$-shell. 
The $e_{\rm max}$=3 space is considered to contain the major 
configurations of such intruder states, but a somewhat larger space 
is needed to stabilize those configurations by coupling to
even higher configurations.

According to the adequacy of the $e_{\rm max}$=3 space, 
the states being discussed are divided into two groups : 
(i) 0$^{+}_{1}$ and 2$^{+}_{1,2}$ states, 
(ii) 0$^{+}_{2}$ and 2$^{+}_{3}$ states. 
For the present interaction and nuclei, 
the $e_{\rm max}$=3 space appears to be adequate for (i), 
whereas is still too small for (ii).
The calculated 2$^{-}_{1}$ and 3$^{-}_{1}$ levels with 
$\beta_{c.m.}\hbar\omega/A =$ 10 MeV are higher than 
experimental ones in Fig.~\ref{be10_excitation_beta}, 
although the deviations are smaller than those for  
0$^{+}_{2}$ and 2$^{+}_{3}$ states.  It is likely that 
the properties of 2$^{-}_{1}$ and 3$^{-}_{1}$ states 
are between the groups (i) and (ii), and should be
investigated also with wider model space.


\begin{figure*}[tb]
\centering
 \includegraphics[width=14cm,clip]{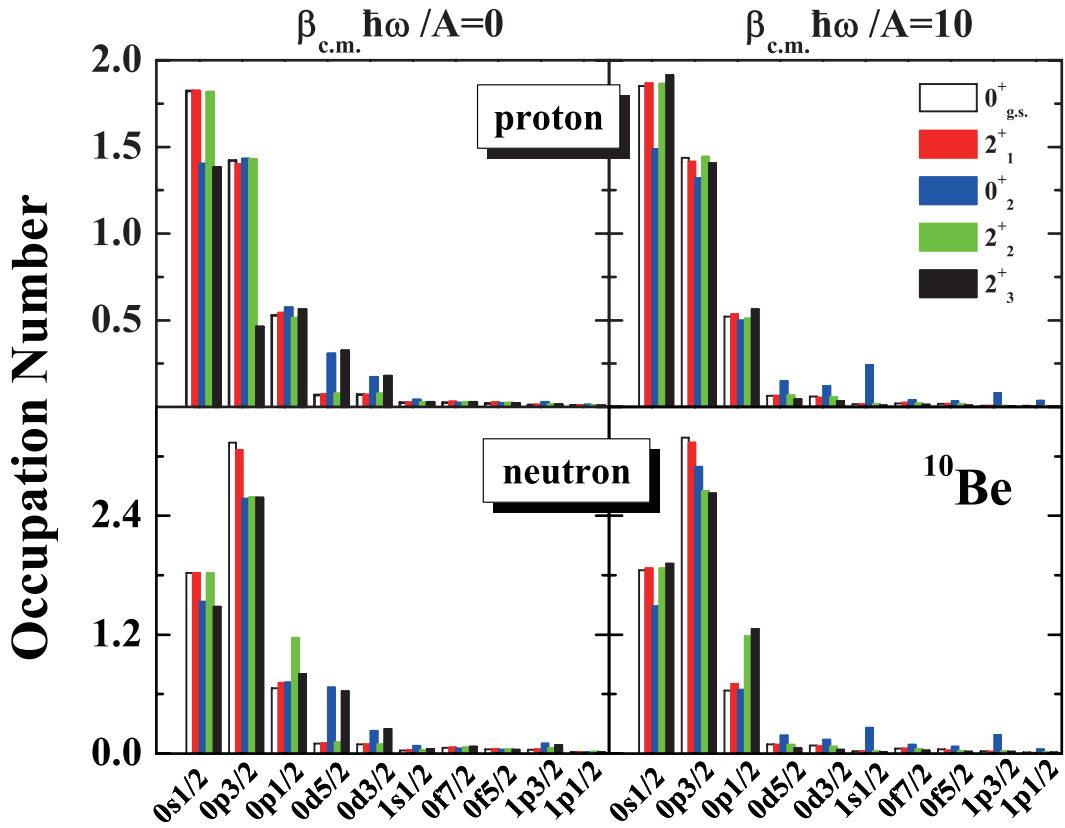}
 \caption{(Color online) The single-particle-orbit occupation numbers of 
 the 0$^+_1$ state (hollow columns), the 2$^+_1$ state (full red
 columns), the 0$^+_2$ state (full blue columns), the 2$^+_2$ state 
 (full green columns) and the 2$^+_3$ state (full black columns) 
 for $^{10}$Be protons (upper panel) and neutrons (lower panel). 
 The wave functions are calculated with 
 $\beta_{c.m.}\hbar\omega/A =$ (a) 0 MeV \, and \, (b) 10 MeV.}
 \label{be10_occup}
\end{figure*}

Figure~\ref{be10_occup} indicates occupation numbers of single-particle
orbits for the ground and some low-lying states of $^{10}$Be.
Left side is for the results without the c.m. motion suppression, 
whereas right side is for those with it (default setup of this work). 
We begin with the group (i), for which left and right sides do not
show much difference.  
This is consistent with almost unchanged level energies of 
2$^{+}_{1,2}$ between the two corresponding calculations, as 
depicted in Fig.~\ref{be10_excitation_beta}.
Figure~\ref{be10_occup} shows that protons and neutrons are mainly 
in the $0s$ and $0p$ orbits.  
The occupation number of the 0$s_{1/2}$ orbit is about 1.8 for
both protons and neutrons.  This value is remarkably constant for
the three states in the group (i), 
and changes very little between proton and neutron.  
This means that the 0$s_{1/2}$ orbit is occupied by $\sim$90 \% 
probability for both proton and neutron.

If the 0$s_{1/2}$ orbit is fully occupied, the $^4$He core is 
ideally formed. 
The present result suggests that the probability of the $^4$He-core
formation is about (0.9)$^4 \sim$ 2/3.  The breaking of the $^4$He 
core is nothing but the polarization of the core, which 
yields effective charges in the shell model with a core.  
In the present calculation, this effect is explicitly treated, 
producing the right amount of B(E2) values as discussed above.
The UCOM transformations act on short-range part of relative-motion 
wave functions.  Electromagnetic operator at long wave length limit 
is then expected to be unaffected to a large extent.  
Thus, we use bare charges and E2 operator, for simplicity.

The occupation number of the $sd$ shell turned out to be 
about 0.7 for protons and neutrons combined, which corresponds 
approximately to the number of nucleons excited from the $^4$He core.
More precise studies on the process of effective charges will be
of much interest. 

For the group (ii) (0$^+_{2}$ and 2$^+_{3}$ states),  
the occupation numbers do change substantially between left and 
right sides of Fig.~\ref{be10_occup}. 
This is consistent with the changes of their excitation 
energies shown in Fig.~\ref{be10_excitation_beta}.

The MCSM results for $^{10}$Be
low-lying spectra can be compared with those of NCSM in
Ref.~\cite{Caurier2002}. Although the NCSM calculation does not use the
same potential, the MCSM calculation shows similar results with the NCSM
calculation for the 2$^{+}_{1}$ and 2$^{+}_{2}$ state. More results of
the NCSM are listed and discussed in Ref.~\cite{Caurier2002} for further
comparison.
The NCSM approach may have some difficulty for 
similar calculations 
because the full $sd$ shell configurations cannot be included at 
8$\hbar \omega$ truncation, for instance. 
In the present MCSM, the $sd$ configurations are fully
included. 


\begin{figure}[tb]
\centering
 \includegraphics[width=8cm,clip]{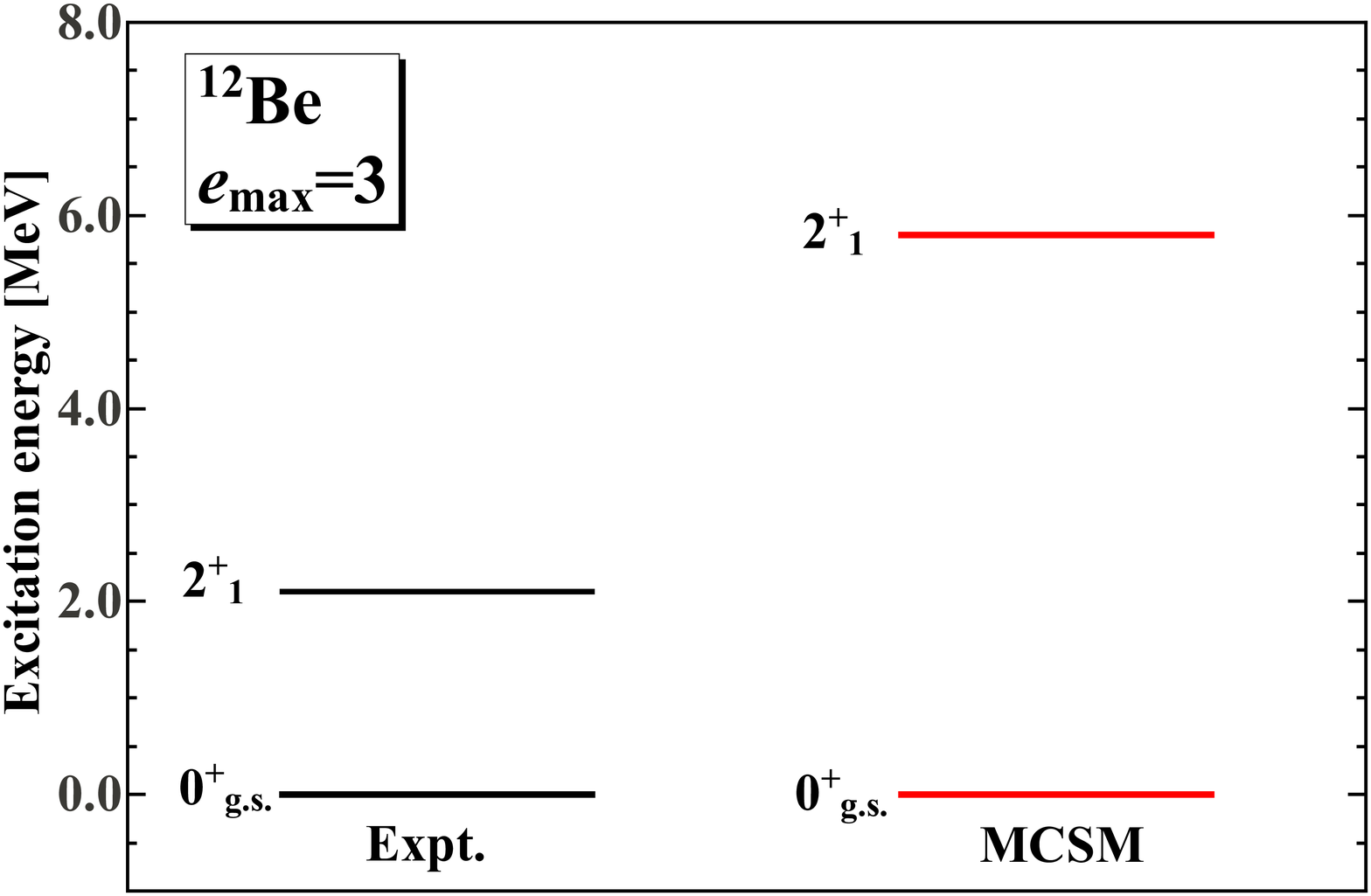}
 \caption{(Color online) The low-lying excitation energy levels of 
$^{12}$Be calculated by the MCSM in $e_{\rm max}$=3 with the UCOM 
potential. The black bars indicate experimental 
data~\cite{Ajzenberg-Selove1988}, while the red bars the MCSM results 
 with $\beta_{c.m.}\hbar\omega/A =$ 0 MeV.
 }
 \label{be12_ex}
\end{figure}

As we know from the experimental data, the 2$^{+}_{1}$ state of
$^{12}$Be has a lower excitation energy than the 2$^{+}_{1}$ state of
$^{10}$Be. This is interpreted as a phenomenon related to the 
evolution of the
magic number in exotic nuclei. However, the present MCSM results in an
$e_{\rm max}$=3 model space do not show this feature as can be
seen in Fig.~\ref{be12_ex}.  The suppression of spurious c.m. motion
may make the discrepancy to experiment larger.  We definitely
need a larger model space, and it is not tractable presently.

\section{\label{sec5}summary}

For the first time, we have applied the no-core MCSM with realistic 
UCOM-transformed interactions to the investigation of structure 
of $^{10}$Be and $^{12}$Be.

We calculate some low-lying states of $^{10}$Be and $^{12}$Be in an 
$e_{\rm max}$=3 model space. The results for the $2^{+}_{1}$ and 
$2^{+}_{2}$ states of $^{10}$Be show a reasonable agreement with 
experimental data. 
We have kept particular attention to the spurious center-of-mass motion, 
and have suppressed by Lawson's method throughout this work.
The MCSM results show negative (positive) quadrupole
moment for the 2$^+_1$ (2$^+_2$) state for $^{10}$Be. 
We have analyzed the
sensitivity of physical observables to center-of-mass contaminations. 
Some states, {\it e.g.}, the 2$^{+}_{1,2}$, are stable against variations of
$\beta_{c.m.}$, but others, e.g., the 1$^-_1$ state, are sensitive to
$\beta_{c.m.}$. In fact, the suppression of spurious c.m. motion 
moves the 1$^-_1$ level by about 4 MeV
to a better agreement to experiment. 
The obtained B(E2) values are 9.29 e$^2\,$fm$^4$ for 
B(E2;$\,2^+_{1}\rightarrow$ 0$^+_{g.s.}$) for $^{10}$Be  
and 9.30 e$^2\,$fm$^4$ for its mirror nucleus $^{10}$C, which are close
to the experimental data with proton bare charge. 
Intrinsic quadrupole moments of the 2$^{+}_{1}$ of $^{10}$Be, 
evaluated from the spectroscopic quadrupole moment and B(E2) value are
similar to each other, suggesting an axially symmetric deformation. 
However, calculated B(E2;$\,2^+_{2}\rightarrow$ 2$^+_{1}$) value is
sizable, being consistent with a modest triaxial deformation.
The triaxial deformation is predicted to be more developed in 
$^{10}$C, providing intriguing issues on the mirror nuclei $^{10}$Be 
and $^{10}$C to be further investigated. 

The MCSM calculation presented here were performed in model spaces up to
$e_{\rm max}$=3. Additional results, e.g. for $^{10}$Be in 
$e_{\rm max}$=2, can be found in Ref.~\cite{liuthesis}. 
Although the ground-state energy of $^{10}$Be is changed by about 10 MeV 
when going from $e_{\rm max}$=2 to $e_{\rm max}$=3, 
other observables turn out to
be more stable already for the $e_{\rm max}$=3 model space. 
Thus, while the use of larger model spaces in the MCSM is certainly 
interesting and resultant changes will be evaluated in future, 
excitation energies and transition matrix elements look converged 
to a certain extent.

\begin{acknowledgments}
  We would like to thank Dr.~Takashi Abe for discussion. One of the authors, Lang Liu, would like to thank the theoretical nuclear structure group in the University of Tokyo and Joint Center for Nuclear Physics in Peking University. This work has been supported by Grants-in-Aid for Scientific Research ((A)20244022) and for Scientific Research on Innovative Areas (20105003) from JSPS. It has also been supported by the CNS-RIKEN joint project for large-scale nuclear structure calculations. This work has also been supported by the Japanese Monbukagakusho Scholarship. R.~Roth acknowledges support from the DFG (SFB 634) and from HIC for FAIR. The calculation was performed partly in Alphleet machine in CNS-RIKEN and partly in T2K super computer in the University of Tokyo.
\end{acknowledgments}

\bibliography{be10}

\end{document}